\documentclass[fleqn,usenatbib]{mnras}

\usepackage{newtxtext,newtxmath,ulem}
\usepackage{tikz}
\usetikzlibrary{decorations.pathreplacing, calc}

\tikzstyle{eq} = [rectangle, text width=27em, rounded corners, minimum height=3em] 
   
\tikzstyle{line} = [draw]
\tikzstyle{data} = [rectangle, draw, fill=white!10,
    text width=3em, text centered, rounded corners, node distance=1.5cm, inner sep=4pt]     
\tikzstyle{eq} = [rectangle, text width=27em, rounded corners, minimum height=3em] 
   
\tikzstyle{prod} = [draw, node distance=1cm,
    minimum height=2em]
\tikzstyle{prob} = [rectangle, draw, fill=green!10,
    text width=7em, text centered, rounded corners, node distance=1cm, inner sep=4pt]
\tikzstyle{det} = [rectangle, draw, fill=blue!10,
    text width=7em, text centered, rounded corners, node distance=1cm, inner sep=4pt] 
    
\tikzstyle{label} = [rectangle, 
    text width=8em, text centered, rounded corners, node distance=2cm, inner sep=4pt]    

\usepackage[T1]{fontenc}

\DeclareRobustCommand{\VAN}[3]{#2}
\let\VANthebibliography\thebibliography
\def\thebibliography{\DeclareRobustCommand{\VAN}[3]{##3}\VANthebibliography}

\usepackage{graphicx}	% Including figure files
\usepackage{amsmath}	% Advanced maths commands
\title[Lifting weak lensing degeneracies]{Lifting weak lensing degeneracies with a field-based likelihood}

\author[N.\ Porqueres et al.]{
Natalia Porqueres,$^{1}$\thanks{n.porqueres@imperial.ac.uk}
Alan Heavens,$^{1}$
Daniel Mortlock,$^{1,2,3}$
and Guilhem Lavaux$^{4}$ 
\\
$^{1}$Imperial Centre for Inference and Cosmology (ICIC) \& Astrophysics group, Imperial College, Blackett Laboratory, \\ Prince Consort Road, London SW7 2AZ, UK\\
$^{2}$Department of Mathematics, Imperial College London, London, SW7 2AZ, UK\\
$^{3}$The Oskar Klein Centre, Department of Astronomy, Stockholm University, Albanova, SE-10691 Stockholm, Sweden\\
$^{4}$CNRS \& Sorbonne Universit\'{e}, UMR7095, Institut d'Astrophysique de Paris, F-75014, Paris, France
}

\date{Accepted 2/11/2021. Received 30/10/2021; in original form 10/08/2021}

\pubyear{2021}

\begin{document}
\label{firstpage}
\pagerange{\pageref{firstpage}--\pageref{lastpage}}
\maketitle

\begin{abstract}
     We present a field-based approach to the analysis of cosmic shear data to infer jointly cosmological parameters and the dark matter distribution. This forward modelling approach samples the cosmological parameters and the initial matter fluctuations, using a physical gravity model to link the primordial fluctuations to the non-linear matter distribution. Cosmological parameters are sampled and updated consistently through the forward model, varying (1) the initial matter power spectrum, (2) the geometry through the distance-redshift relationship, and (3) the growth of structure and light-cone effects. Our approach extracts more information from the data than methods based on two-point statistics. We find that this field-based approach lifts the strong degeneracy between the cosmological matter density, $\Omega_\mathrm{m}$, and the fluctuation amplitude, $\sigma_8$, providing tight constraints on these parameters from weak lensing data alone. In the simulated four-bin tomographic experiment we consider, the field-based likelihood yields marginal uncertainties on $\sigma_8$ and $\Omega_\mathrm{m}$ that are, respectively, a factor of 3 and 5 smaller than those from a two-point power spectrum analysis applied to the same underlying data.
     
\end{abstract}

\begin{keywords}
    cosmology:large-scale structure of Universe -- methods:data analysis -- weak gravitational lensing
\end{keywords}

%%%%%%%%%%%%%%%%%%%%%%%%%%%%%%%%%%%%%%%%%%%%%%%%%%

%%%%%%%%%%%%%%%%% BODY OF PAPER %%%%%%%%%%%%%%%%%%

\section{Introduction}

Weak gravitational lensing by large-scale structure modifies the shape of galaxy images through a process known as cosmic shear. It is a powerful probe of cosmology as it is sensitive to the geometry of the Universe and the growth of cosmic structures, and is not dependent on a knowledge of galaxy bias. Cosmic shear data have provided useful constraints on the key cosmological parameters that describe the dark sector of the Universe \citep{Troxel2018, Hikage19, Hamana2020, Asgari2021, Amon2021, Secco2021}, especially the amplitude of matter fluctuations. These cosmological constraints are usually inferred from the analysis of two-point statistics, either the lensing power spectrum \citep{Hikage19, Asgari2021} or the shear two-point correlation functions \citep{Hildebrandt2017,Troxel2018,Hamana2020, Asgari2021, Amon2021, Secco2021}. These standard methods based on the two-point summary statistics are sub-optimal for the non-Gaussian shear field and discard phase information. This has led to the development of several alternative inference techniques such as third- and higher-order statistics \citep{Takada2002Third, Pen2003, Bernardeau2005Third, Jarvis2004, Kilbinger2005Third, Semboloni2011Third, Waerbeke2013, Fu2014Third, Jung2021}, lensing peak count statistics \citep{Jain2000, Dietrich2010, Maturi2011, Lin2015, Liu2015Peaks, Kacprzak2016Peaks, Petri2013, Peel2017, Fluri2018Peaks, Martinet2018Peaks, Shan2018Peaks, Harnois2021Peaks} the lensing probability distribution function \citep{Boyle2021PDF, Martinet2021PDF}, shear clipping \citep{Giblin2018}, and machine learning approaches \citep{Gupta2018ML, Fluri2018ML, Jeffrey2021ML, Ribli2019ML}. All these methods have the potential to improve the cosmology constraints, but they still rely on summary statistics that require an assumption of their sampling distribution, usually assumed to be Gaussian, which in turn requires a covariance matrix, which is difficult to compute reliably. One way to avoid the use of summary statistics and the problems associated with them is through data assimilation methods, in which the observations are incorporated into a forward model. Some forward modelling approaches have been developed for cosmic shear analysis \citep{Boehm17, Alsing16, Alsing17, BORG-WL, Fiedorowicz2021}. These methods differ in the quantities they sample and their prior assumptions. We have developed one such method by extending the Bayesian Origin Reconstruction from Galaxies \citep[BORG,][]{BORG,BORG-3} formalism to incorporate weak lensing as {\sc BORG-WL} \citep{BORG-WL}. {\sc BORG-WL} differs from other forward modelling approaches in using a physical model of structure formation to link the primordial fluctuations to the evolved matter distribution, allowing us to sample the initial conditions, which follow simple Gaussian statistics. This data model, more complex and more complete, provides a better link between theory and data than using Gaussian (lognormal) priors for the shear (density) field.

Here we present an extended version of {\sc BORG-WL} in which the cosmological parameters are sampled and updated consistently in the forward model. We vary the primordial matter power spectrum, the geometry of the universe (through the distance-redshift relation) and the growth rate of structures (through the dynamical model and its light-cone). We compared the results of {\sc BORG-WL} to a standard analysis based on lensing power spectra applied to the same simulated data. 

This paper is organised as follows. In Section \ref{sec:data_model}, we describe the data model. In Section \ref{sec:method}, we present the inference method used in this work, {\sc BORG-WL}. Section \ref{sec:mock} describes the simulated data employed to test and validate the method. The results are presented in Section \ref{sec:results}, which includes the cosmology constraints from {\sc BORG-WL} and a comparison to the standard analysis based on lensing power spectra. We also discuss the efficiency and convergence of the method and present the results of the validation tests. Section \ref{sec:conclusions} summarises the results. 

\section{Data model}
\label{sec:data_model}

The effect of weak gravitational lensing on a source can be described by the shear, $\gamma$, which indicates the distortion in the shape of the image, and the convergence field, $\kappa$, which describes the  variation in angular size. In the flat-sky approximation, which we assume throughout, the Fourier transforms of these two fields, $\tilde{\gamma}$ and $\tilde{\kappa}$, are related by
\begin{align}
    \Tilde{\gamma}(\boldsymbol{\ell}) = \frac{(\ell_1 + i \ell_2)^2}{\ell^2} \Tilde{\kappa}(\boldsymbol{\ell}),
\label{eq:shear}
\end{align}
where $\boldsymbol{\ell} = (\ell_1, \ell_2)$ is the wavevector and the shear is written as a complex quantity. The convergence field is connected to the dark matter distribution by integrating the fractional overdensity along the line-of-sight using the lensing weights to give \citep{KilbingerReview}
\begin{equation}
    \kappa(\boldsymbol{\vartheta}) = \frac{3 H_0^2 \Omega_\mathrm{m}}{2 c^2} \int^{r_\mathrm{lim}}_0 \frac{rdr}{a(r)} q(r) \delta^{\rm f}(r\boldsymbol{\vartheta}, r),
\label{eq:kappa}
\end{equation}
where $\boldsymbol{\vartheta}$ is the coordinate on the sky, $r$ is the comoving distance, $r_\mathrm{lim}$ is the limiting comoving distance of the galaxy sample, $\delta^{\rm f}$ is the dark matter overdensity at a scale factor $a$ and 
\begin{equation}
    q(r) = \int^{r_\mathrm{lim}}_r dr' n(r') \frac{r'-r}{r'},
\end{equation}
with $n(r)$ being the redshift distribution of galaxy sources. We assume a spatially flat universe throughout. 

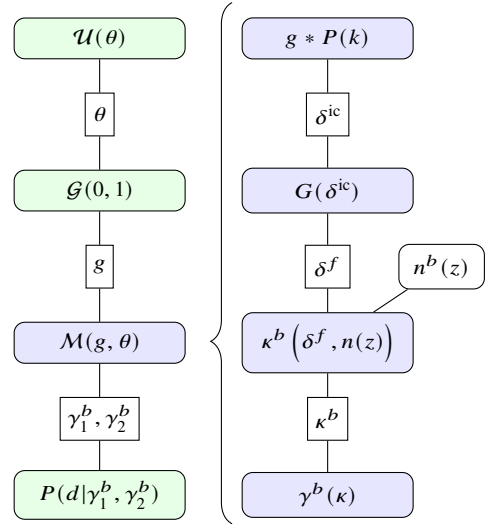
\begin{figure}
\centering
\begin{tikzpicture}[node distance = 0.8cm, auto]
    
    \node[prob] (prior) {$\mathcal{U}(\theta)$};
    \node[prod, below of=prior] (S8) {$\theta$}; 
    \node[prob, below of=S8] (prior_ic) {$\mathcal{G}(0,1)$};
    \node[prod, below of=prior_ic] (ic) {$g$};
    \node[det, below of=ic] (fwd) {$\mathcal{M}(g,\theta)$};
	\node[prod, below of=fwd] (shear) {$\gamma^b_1, \gamma^b_2$};
	\node[prob, below of=shear] (likelihood) {$P (d|\gamma^b_1, \gamma^b_2)$};
    
    \begin{scope}
    \node[det, right of=prior, node distance=3cm] (conv) {$g*P(k)$};
    \node[prod, below of=conv] (convolved) {$\delta^\mathrm{ic}$};
    \node[det, below of=convolved] (grav) {$G(\delta^\mathrm{ic})$};
	\node[prod, below of=grav] (delta) {$\delta^f$};
	\node[det, below of=delta] (kappam) {$\kappa^b\left(\delta^f, n(z)\right)$}; 
	\node[prod,below of=kappam] (kappa) {$\kappa^b$}; 
	\node[det, below of=kappa] (shearm) {$\gamma^b(\kappa)$};
	\node[data, right of=delta](sources) {$n^b(z)$}; 
	\end{scope}
	
    \draw [decorate,decoration={brace,amplitude=8pt,aspect=0.35},xshift=-50pt,yshift=-200pt] (3.5,0.65) -- (3.5,7.5)node [black,xshift=9pt]  {};

    % Draw edges
    \path [line] (prior) -- (S8);
    \path [line] (S8) -- (prior_ic);
    \path [line] (prior_ic) -- (ic);
    \path [line] (ic) -- (fwd);
    \path [line] (fwd) -- (shear);
    \path [line] (sources) -- (kappam);
    \path [line] (shear) -- (likelihood);
    \path [line] (conv) -- (convolved);
    \path [line] (convolved) -- (grav);
    \path [line] (grav) -- (delta);
    \path [line] (delta) -- (kappam);
    \path [line] (kappam) -- (kappa);
    \path [line] (kappa) -- (shearm);
\end{tikzpicture}
\caption{Representation of the Bayesian hierarchical model. Blue boxes indicate deterministic functions and green boxes represent probability distributions. $\theta$ are the cosmological parameters, $g$ is a Gaussian field with zero mean and unit variance, $\mathcal{M}(g,\theta)$ is the forward model, which consists of a convolution of the Gaussian field with the initial matter power spectrum $P(k)$ to obtain the initial conditions $\delta^\mathrm{ic}$, a gravity and structure formation model $G(\delta^\mathrm{ic})$ and the data model to obtain the shear $\gamma^b$ for each tomographic bin $b$. $P (d|\gamma^b_1, \gamma^b_2)$ is the likelihood with $d$ being the data.}
\label{fig:BHM}
\end{figure}  

These fields can be seen as latent parameters of a Bayesian hierarchical model as represented in Fig. \ref{fig:BHM}. The starting point is the cosmological parameters $\boldsymbol{\theta}$. Given a cosmology, we generate primordial fluctuations, $\delta^\mathrm{ic}$, as Gaussian random fields in a rectangular cuboid of voxels, with a covariance matrix corresponding to the initial matter power spectrum. To compute the initial power spectrum we used the prescription of \citet{EH98, EH99}, which includes baryonic acoustic oscillations. The primordial fluctuations are linked to the evolved large-scale structures by the non-linear gravity model $\delta^\mathrm{f}=G(\delta^\mathrm{ic})$, which describes the growth of cosmic structures and accounts for light-cone effects. We then use the evolved density field, $\delta^\mathrm{f}$, to generate the convergence field $\kappa$. In our discrete implementation  and using the Born approximation, the radial line-of-sight integral in Equation~\ref{eq:kappa} is approximated by a sum over voxels as
\begin{equation}
    \kappa^b_{mn} =\frac{3 H_0^2 \Omega_\mathrm{m}}{2 c^2} \sum\limits_{j=0}^{N}  \delta^\mathrm{f}_{mnj} \left[\sum\limits_{s=j}^{N} \frac{(r_s - r_j)}{r_s} n^b(r_s) \Delta r_s \right] \frac{r_j \Delta r_j}{a_j},
\end{equation}
where the index $b$ indicates the tomographic bin, and the sub-indices $m$ and $n$ label the pixel on the sky, which is chosen to be large enough to contain many sources. The index $j$ labels the voxels along the line-of-sight at a comoving distance $r_j$. $N$ is the total number of voxels along the line of sight, which are found using a ray tracer. $\Delta r_j$ is the length of the line of sight segment inside the voxel $j$, and $\delta_f$ is the three-dimensional dark matter distribution. The comoving radial distance $r_s$ indicates the distance to the source. The redshift distribution of sources is given by $n^b(z_s)$ for each tomographic bin. Having evaluated $\kappa_{mn}^b$ in this way we then use equation \ref{eq:shear} to obtain the predicted shear field. 

In this work, we do not include the important effects of intrinsic alignments or baryon feedback in our forward model, and we leave these to future work, where we will also include the uncertainty in the redshift distributions of the sources. These effects can be included in the inference as associated nuisance parameters.

\section{Method}
\label{sec:method}

In this work, we extend the Bayesian hierarchical model presented in \cite{BORG-WL}, {\sc BORG-WL}, to infer the cosmological parameters jointly with the density field. Here we briefly describe the method and indicate the changes we have implemented to simultaneously constrain the cosmology and density field. A more detailed description of the general {\sc BORG} framework can be found in \citet{HADES, BORG, BORG-3}.

The {\sc BORG-WL} method is a Bayesian framework to infer jointly the cosmology and the linear and non-linear matter distributions from pixelated tomographic shear fields. The underlying idea is to infer the initial conditions and the cosmological parameters from the data, using a full gravitational structure formation model. For that, the {\sc BORG} framework employs a non-linear gravity model for structure growth, either based on perturbation theory or N-body simulations.  This dynamical model connects the primordial fluctuations to the evolved large-scale structures, allowing us to sample the primordial fluctuations, which follow simple Gaussian statistics. The {\sc BORG} framework, therefore, uses a Gaussian prior for the initial conditions at an initial scale factor $a \approx 10^{-3}$. This Gaussian prior has zero mean and a covariance matrix corresponding to the initial matter power spectrum. In this extension of {\sc BORG-WL} for cosmology inference, we vary the initial matter power spectrum according to the value of the cosmological parameters given by the sample, using the transfer function of \cite{EH98, EH99}.

We have implemented a ray tracer to identify the voxels corresponding to each line of sight. With this ray tracer, we drop the distant observer approximation from our previous work and integrate radially from an observation point using the Born approximation. The line-of-sight integration depends on the cosmology through the distance-redshift relation, which we vary when sampling the cosmological parameters. The cosmology sampler, therefore, varies the geometry of the universe consistently with the cosmology. 

Another element of the model that is sensitive to cosmology is the structure formation prescription. The dynamics and light-cone effects also vary consistently with the cosmological parameters. {\sc BORG-WL}, therefore, samples and updates the cosmology consistently throughout the forward model by changing all the quantities that depend on the cosmological parameters. This consistent cosmology sampler is a powerful new addition to the {\sc BORG} framework. Previously, \cite{Altair} presented a proof-of-concept to sample cosmological parameters with {\sc BORG} from galaxy clustering data. However, their cosmology sampler did not update the forward model consistently, and the initial power spectrum was fixed. In contrast, our model is consistent, and its sensitivity to the cosmology is three-fold: through the geometry of the problem, the growth of structures, and the initial matter power spectrum. This allows us to provide joint constraints on the cosmological parameters and the density field. 

Our method proceeds as follows. Given values for the cosmological parameters, we generate a random realisation of the three-dimensional primordial fluctuations as a Gaussian field with a covariance matrix corresponding to the initial matter power spectrum. The structure formation model then evolves these initial conditions to a non-linear realisation of the dark matter distribution, correctly accounting for light-cone effects. By integrating the 3D dark matter field along the lines of sight and applying the lensing data model described in section \ref{sec:data_model}, {\sc BORG-WL} predicts the two components of the pixelated tomographic shear fields on the flat sky, $\gamma_{1,mn}^b$ and $\gamma_{2,mn}^b$, where $b$ denotes the tomographic bin and $m$ and $n$ are the sky pixel indices.

The measured shear in each pixel and bin, $\hat{\gamma}_{1,mn}^b$ and $\hat{\gamma}_{2,mn}^b$, will differ from the predicted values due to a combination of the galaxy shape measurement errors and range of intrinsic source shapes of the galaxies in the voxel.  This is encoded in a likelihood of the form $P(\hat{\gamma}_{1,mn}^b, \hat{\gamma}_{2,mn}^b | \gamma_{1,mn}^b, \gamma_{2,mn}^b)$.  Our numerical implementation does not rely on this likelihood having any specific mathematical form, and in particular can handle shear noise that varies with both sky pixel and tomographic bin.  For the simulations used here we assume that the observations are characterised by shape noise with variance $\sigma_\epsilon^2$, and that the associated shear uncertainty is hence given by $\sigma_b = \sigma_\epsilon / \sqrt{N_b}$, where $N_b$ is number of sources per pixel in tomographic bin $b$ (assumed here to be independent of sky pixel).  For sufficiently high $N_b$ the voxel likelihood can be approximated as being Gaussian with variance $\sigma_b^2$, which leads to a full log-likelihood of the form 
\begin{align}
    \log \mathcal{L} & =
    \sum\limits_{b}\sum\limits_{mn} \log\left[P(\hat{\gamma}_{1,mn}^b, \hat{\gamma}_{2,mn}^b | \gamma_{1,mn}^b, \gamma_{2,mn}^b)\right]
    \nonumber \\
     & =
         - \sum\limits_{b}\sum\limits_{mn} \frac{\left(\hat{\gamma}^b_{1,mn}-\gamma^b_{1,mn}\right)^2 + \left(\hat{\gamma}^b_{2,mn}-\gamma^b_{2,mn}\right)^2}{2\sigma_b^2},
\end{align}
where normalising constants have been ignored and the dependence of the predicted shears on the underlying parameters is left implicit.  This simple Gaussian likelihood is sufficient in this first demonstration of this method, but a more realistic form could be used when analysing a real dataset, or the hierarchical model could be extended to sample the shear as well \citep{Schneider15}.

\begin{figure}
    \includegraphics[width=\columnwidth]{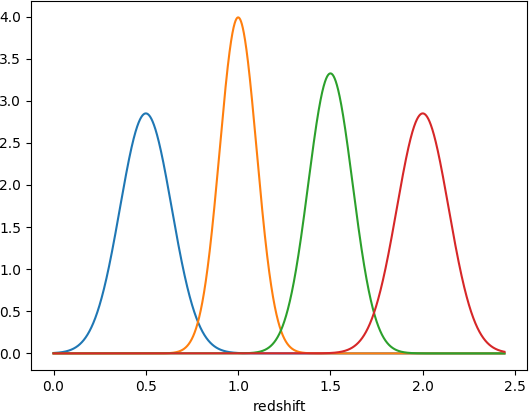}
    \caption{Redshift distributions of sources for each tomographic bin in this analysis.}
    \label{fig:tomobin}
\end{figure}

\subsection{Sampling scheme}

Sampling the density field implies that the amplitude of the density fluctuations in each voxel is a target parameter of the problem. This results in a high-dimensional problem, with typically $10^7$ parameters, that requires advanced statistical techniques to sample from the posterior distribution. We have split the problem in two sampling blocks following a Gibbs sampling scheme, with the density and the cosmological parameters sampled alternately according to 
\begin{align}
    \boldsymbol{\delta}^\mathrm{ic} \curvearrowleft
    P(\boldsymbol{\delta}^\mathrm{ic}|\boldsymbol{\theta}, \boldsymbol{d}), \\
    \boldsymbol{\theta} \curvearrowleft P(\boldsymbol{\theta} | \boldsymbol{\delta}^\mathrm{ic}, \boldsymbol{d}).
\end{align} 
To deal with the high number of parameters in the density field, {\sc BORG} employs a Hamiltonian Monte Carlo sampler \citep[HMC,][]{Neal2011}, which uses the information in the gradients to navigate the parameter space. To sample the cosmology, we use a slice sampler. In both steps of the Gibbs sampling, the cosmology is consistent throughout the forward model. 

In a follow-up work, we intend to include systematics modelling in our forward model and add the associated nuisance parameters to the inference. Sampling these few additional parameters will have a minimal extra cost and allow us to propagate their uncertainties automatically. 

\subsection{Setup for this work}

In this work, we used Lagrangian perturbation theory (LPT) as our model of gravitational clustering and structure formation. At the resolution we use (14$h^{-1}$~Mpc), the LPT model does not show significant deviations from particle-mesh simulations \citep{Tassev2013}. For future applications of this method at high resolution, we will use a fully non-linear particle-mesh such as the one described in \cite{Jasche18BorgPM} to have an accurate description of the matter density at smaller scales.

\begin{figure}
    \includegraphics[width=\columnwidth]{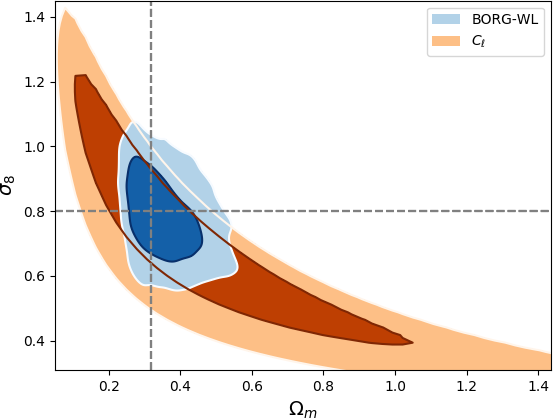}
    \caption{Comparison of $\Omega_\mathrm{m}$-$\sigma_8$ constraints from our method {\sc BORG-WL} (blue) and the standard analysis based on lensing tomographic power spectra (orange). The contours correspond to the 68.3\% and 95.5\% highest posterior density credible regions. Both posteriors are obtained by applying both methods to the same simulated shear data, with 4 tomographic bins and 30 galaxies per square arcmin. The dashed lines indicate the true values of the parameters.}
    \label{fig:borg_cls}
\end{figure}

\begin{figure*}
    \includegraphics[width=0.8\hsize,clip=true]{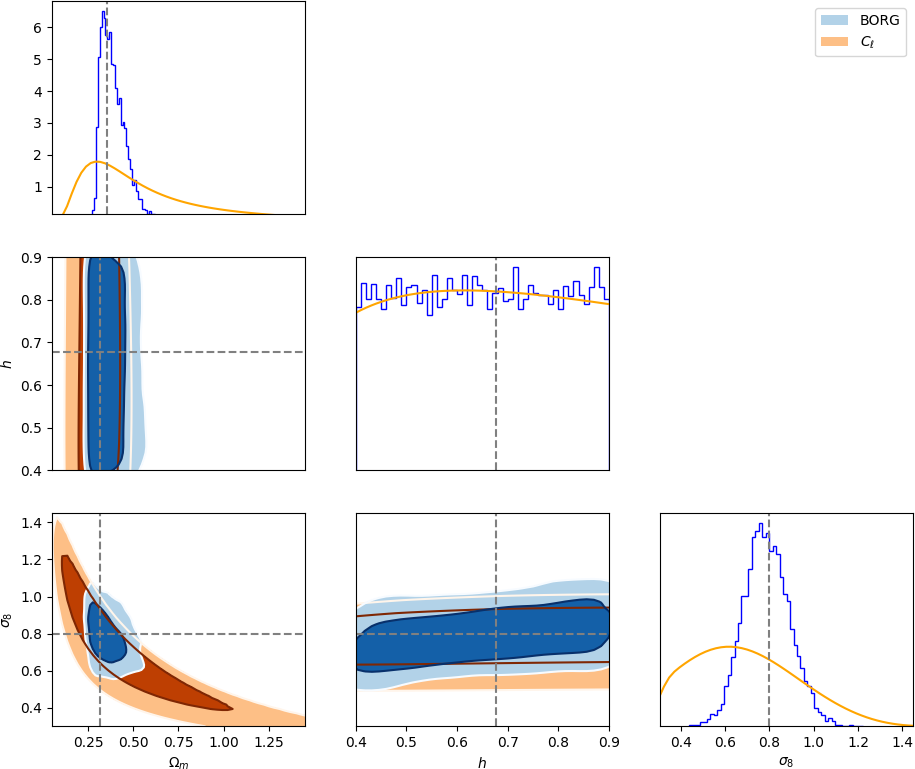}
    \caption{Posterior distribution of the sampled cosmological parameters from the {\sc BORG-WL} analysis (blue) and the standard analysis based on the lensing power spectra (orange). The dashed lines indicate the ground truth. }
    \label{fig:corner}
\end{figure*}

Here we focus on constraining three cosmological parameters $\Omega_\mathrm{m}$, $h$ and $\sigma_8$, jointly with the density. We consider a flat Universe, so $\Omega_\Lambda = 1 - \Omega_\mathrm{m}$ can be treated as a derived parameter. We use the following uniform priors for the cosmological parameters: $\Omega_\mathrm{m} \mathtt{\sim} {\cal U}[0.1, 0.8]$; $h \mathtt{\sim} {\cal U}[0.4, 0.9]$; and $\sigma_8 \mathtt{\sim} {\cal U}[0.1,4.0]$. To improve the efficiency of the sampler, we perform a coordinate transformation and sample $\boldsymbol{\theta}' = (\Omega_\mathrm{m}, h, S_8\equiv\sigma_8\left(\Omega_\mathrm{m}/0.3\right)^{0.5})$ instead of $\boldsymbol{\theta} = (\Omega_\mathrm{m}, h, \sigma_8)$. This coordinate transformation reduces the correlation length of the sampler by a factor of 4, making the inference more efficient.

\section{Mock data}
    \label{sec:mock}
    
    To test our method, we generated mock data using the forward model illustrated in Fig. \ref{fig:BHM}. 
    
    To generate the mock data, we assumed a standard $\Lambda$CDM cosmology with the following parameters: $\Omega_\mathrm{m} = 0.3175,\ \Omega_\Lambda = 0.6825,\ \Omega_b = 0.049,\ h=0.677,\ \sigma_8= 0.8,\ n_s = 0.9624$ with ${\rm H}_0=100 h$~km~s$^{-1}$~Mpc$^{-1}$. Then, we generated Gaussian initial conditions on a Cartesian grid of size ($1 \times 1 \times 4$) $h^{-1}$~Gpc, with $64\times 64 \times 128$ voxels. 
    We evolved the primordial fluctuations via LPT, accounting for light-cone effects. To obtain the evolved density fields, we used a cloud-in-cell scheme to estimate the density on the Cartesian grid from simulated particles. We generated tomographic shear fields by applying the lensing data model in Sec. \ref{sec:data_model} and using the redshift distribution of sources shown in Fig. \ref{fig:tomobin}. The angular scale of the voxels for the different tomographic bins is 40 arcmin, 23 arcmin, 17 arcmin and 15 arcmin. Finally, we added Gaussian pixel noise with a variance that corresponds to $30$ galaxies per square arcmin, as expected for the {\it Euclid} survey \citep{EuclidStudyReport}, and with an uncertainty on intrinsic ellipticity $\sigma_\epsilon = 0.3$ \citep[][]{KilbingerReview}. In this experiment, we set $\sigma_\epsilon$ as the variance of a single component of ellipticity. The sources are then distributed uniformly over the four tomographic bins. The resulting noise level for the farthest tomographic bin is $N=0.013$, and the data is noise-dominated at all scales.
    
\section{Results}
\label{sec:results}

\begin{figure*}
    \includegraphics[width=\hsize]{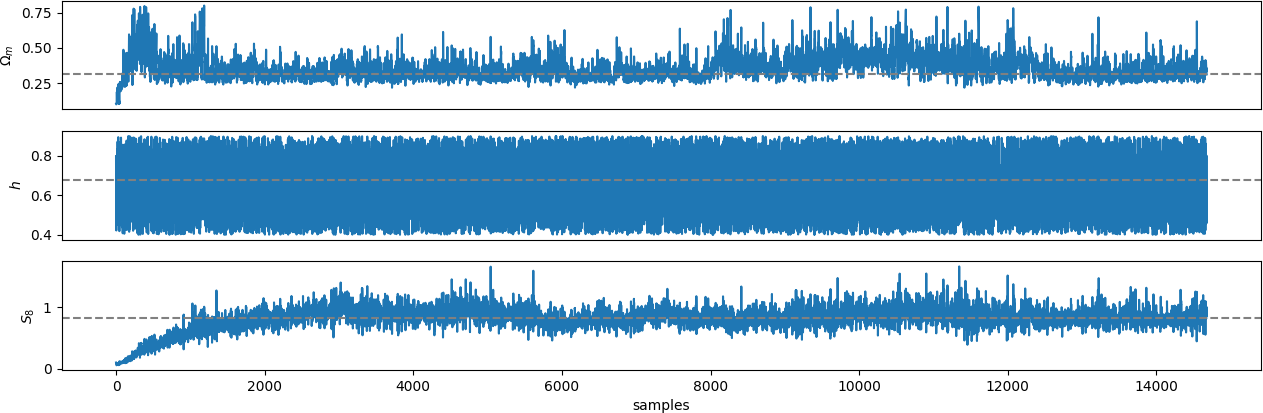}
    \caption{Trace plot for the cosmology sampler. The burn-in phase can be identified as the drift of the parameters in the first $\sim 2500$ samples, in particular $S_8$. The dashed lines indicate the true value of the parameters.}
    \label{fig:trace}
\end{figure*}

\begin{figure}
    \includegraphics[width=\columnwidth]{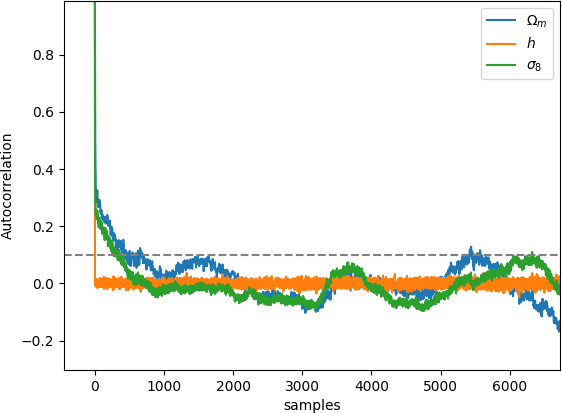}
    \caption{Auto-correlation of the cosmological parameters as a function of the sample number in the Markov chain. The correlation length of the sampler can be estimated as the point when the correlation drops below 0.1.}
    \label{fig:corr}
\end{figure}

\subsection{Cosmology constraints and comparison to standard analysis}
\label{sub:comparison_results}

Here we present the mock-constraints on cosmology obtained by applying {\sc BORG-WL} to the simulated data described in Sec. \ref{sec:mock}. We also compare our results to one of the common techniques for analysing shear data, i.e. an analysis of the lensing power spectra assuming a Gaussian likelihood with a fixed covariance matrix.

To compare the results of {\sc BORG-WL} to the standard 2-point statistics approach, we focus on the angular power spectra $C^{ab}_\ell$, defined implicitly by 
\begin{align}
    \langle \kappa^a(\ell) \kappa^{b*}(\ell') \rangle = (2\pi)^2 \delta_D(\ell - \ell') C^{ab}_\ell,
\end{align}
where $a$ and $b$ label the tomographic bins. Following the common approach in lensing analyses, we assume that the sampling distribution of the $C^{ab}_\ell$ is Gaussian, 
\begin{align}
    \log\mathcal{L} = - \frac{1}{2} \left[\boldsymbol{d} - \boldsymbol{m}(\boldsymbol{\theta})\right]^{\rmn T} \boldsymbol{\mathrm{C}}^{-1}\left[\boldsymbol{d} - \boldsymbol{m}(\boldsymbol{\theta})\right] + \mathrm{constant},
    \label{eq:gaussian_summaries}
\end{align}
where the data vector $\boldsymbol{d}$ contains the auto- and cross-power spectra of the tomographic bins. $\boldsymbol{\mathrm{C}}$ is the covariance matrix, which is assumed to be fixed and computed at a fiducial cosmology $(\Omega_\mathrm{m}^f, \sigma_8^f, h^f) = (0.3, 0.8, 0.67)$. To compute the covariance matrix, we used the Gaussian approximation in \cite{KilbingerReview},  
\begin{equation}
    {\rm Cov}(C_\ell, C_\ell') = \frac{1}{f_\mathrm{sky} \, \ell \, \Delta \ell} \left( C_\ell + N_\epsilon \right)^2 \delta_{\ell \ell'}
    \label{eq:cov}
\end{equation}
where $f_\mathrm{sky}$ is the fraction of the sky observed by the survey, $\Delta \ell$ is the width of the $\ell$-bin, $\delta_{\ell\ell'}$ is a Kronecker delta, which makes the covariance diagonal, and $N_\epsilon$ is the noise power given by $\sigma^2_\epsilon / 2n_g$, with $\sigma_\epsilon$ being the intrinsic dispersion of galaxy ellipticities and $n_g$ being the mean number density of source galaxies on the sky. For this analysis, we use broad uniform priors for $\Omega_\mathrm{m}$, $\sigma_8$ and $h$ on the intervals $[0.05, 1.45]$, $[0.3, 1.45]$ and $[0.4, 0.9]$, sufficiently broad that the posterior is unaffected. To compute the model prediction $\boldsymbol{m}(\boldsymbol{\theta})$, we use the Core Cosmology Library \citep{CCL}.

\begin{figure}
    \includegraphics[width=\columnwidth]{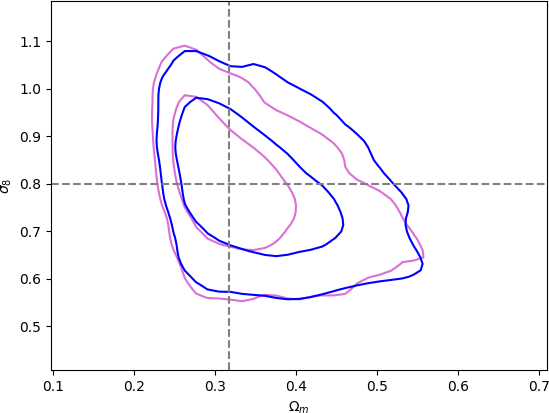}
    \caption{Contours of 68.3\% and 95.5\% highest posterior density credible regions for two chains with different starting points. As can noted here, both chains converge qualitatively to the same distribution.}
    \label{fig:convergence}
\end{figure}

\begin{figure*}
    \includegraphics[width=\hsize,clip=true]{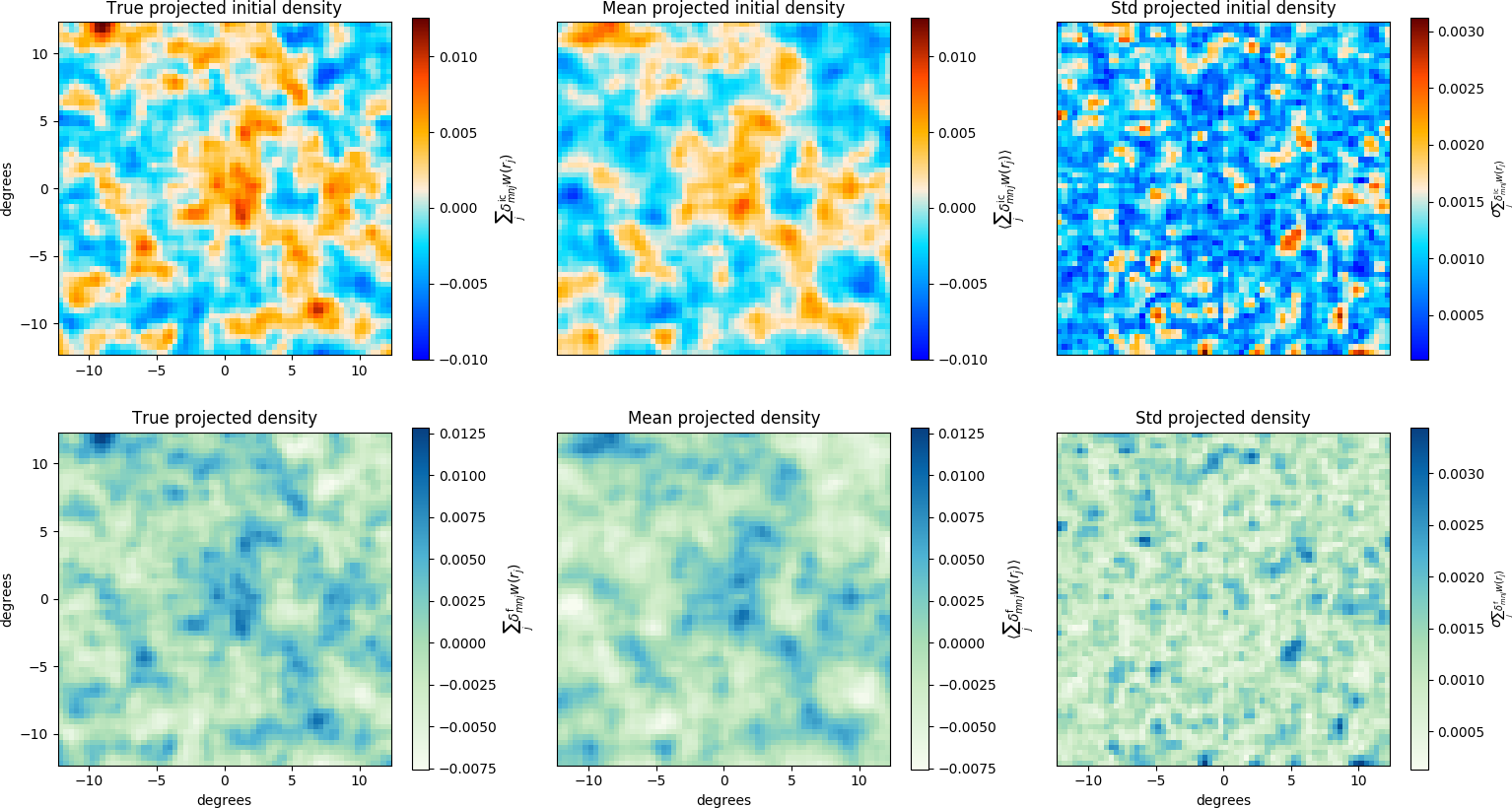}
    \caption{Projection of the initial (upper panels) and evolved (bottom panels) density fields. The first column shows the true fields that we used to generate the data, the second column shows the ensemble mean and the third column shows the standard deviation of the fields. The mean and standard deviation are estimated from 500 MCMC samples.}
    \label{fig:panels_density}
\end{figure*}

\begin{figure}
    \includegraphics[width=0.9\columnwidth]{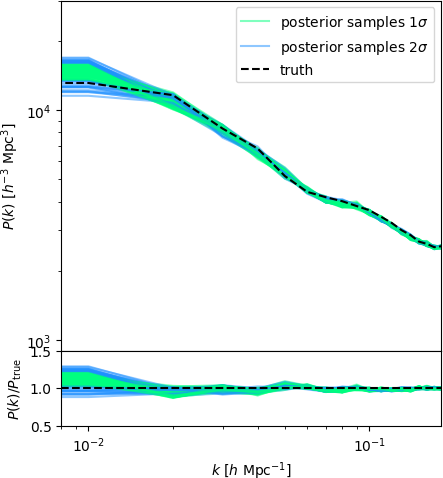}
    \caption{Posterior initial matter power spectrum, measured from the posterior samples of the three-dimensional initial conditions. The input power spectrum is shown as the dashed curve.}
    \label{fig:pk}
\end{figure}

\begin{figure*}
    \includegraphics[width=\hsize,clip=true]{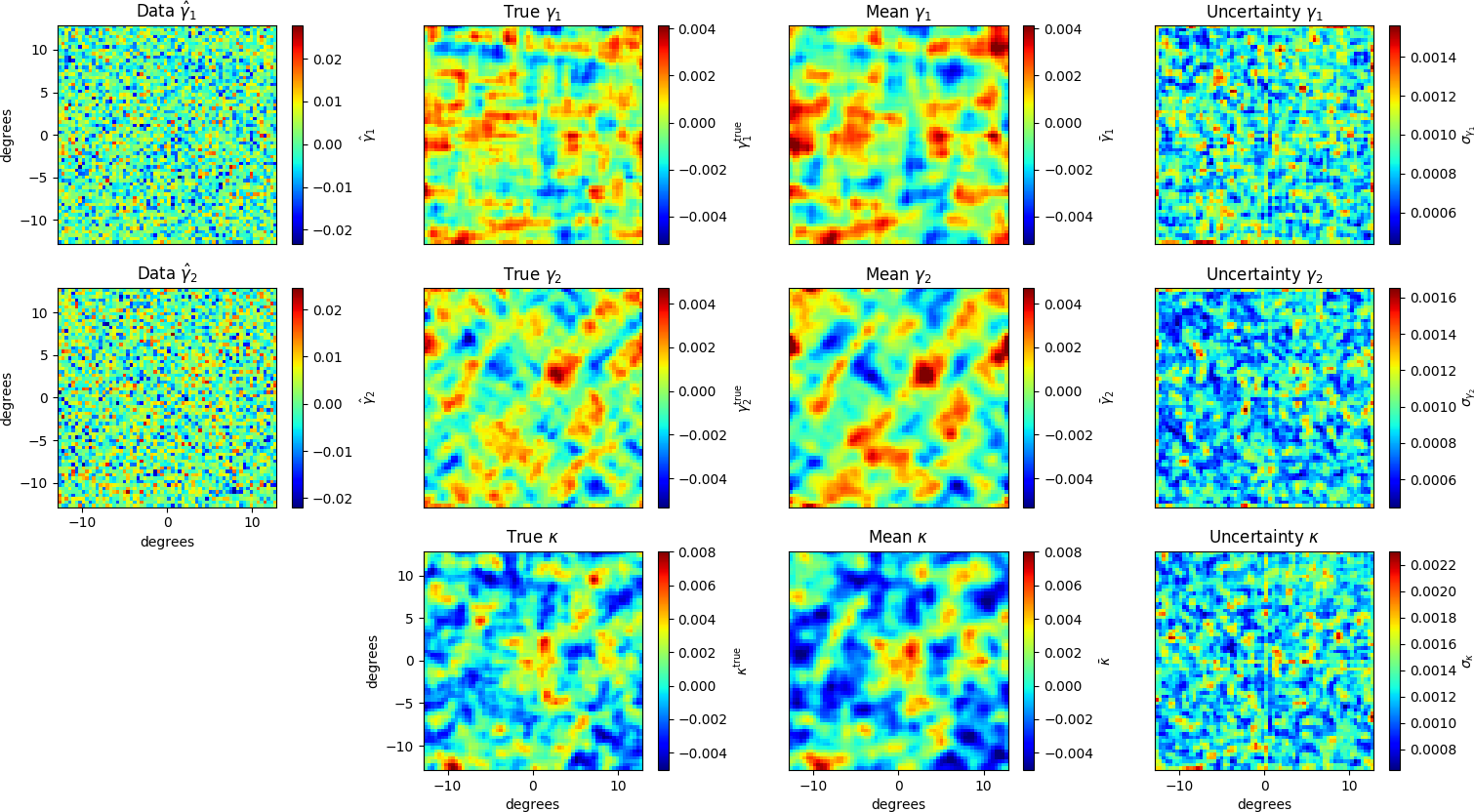}
    \caption{Posterior predicted maps for one tomographic bin. The first column shows the noisy shear mock data, the second column shows the true shear and convergence fields, and the third and fourth columns are the mean and standard deviation computed from 500  posterior-predicted fields.}
    \label{fig:panels_ppt}
\end{figure*}

Figure \ref{fig:borg_cls} shows the 68.3\% and 95.5\%  credible regions for $\Omega_\mathrm{m}$ and $\sigma_8$ obtained with {\sc BORG-WL}. For comparison, we added the constraints from the standard power spectra analysis on the same underlying data, showing that {\sc BORG-WL} lifts the weak lensing degeneracy and yields much tighter constraints on these parameters than the standard analysis based on two-point summary statistics. We note that the size of the posterior distribution is significantly smaller than the adopted size for the prior distributions for $\Omega_\mathrm{m}$ and $\sigma_8$. We conclude that the constraints on these parameters should not be affected by the specific choice of prior. Figure \ref{fig:borg_cls} is the main result of this paper.

 Figure \ref{fig:corner} shows the full corner plot including $h$ and the marginal distributions.  We computed the {\sc BORG-WL} posteriors from 10000 samples after the burn-in phase, which is determined by examining the trace plots of the parameters in Fig. \ref{fig:trace}. These results show that {\sc BORG-WL}  recovers the correct value of $\Omega_\mathrm{m}$ and $\sigma_8$. Since the lensing data is not informative about the value of $h$ \citep{Hall2021}, this parameter is not well constrained, and the marginal distribution corresponds closely to the prior. For comparison, we added the constraints from the standard analysis based on the lensing power spectra. The corresponding posterior distributions are evaluated in grids of  $50 \times 50$ points. The comparison of the marginal distributions shows that {\sc BORG-WL} provides far more precise cosmology constraints than the standard method, improving the error bars of $\Omega_\mathrm{m}$ by a factor 5 and $\sigma_8$ by a factor 3 (see Table \ref{Table1}). The ratio of the constraints on $S_8$ from $C_\ell$ vs BORG-WL is 1.2. The improvement on the constraints of $S_8$ is small compared to $\Omega_m$ and $\sigma_8$ because $S_8$ is a combination of cosmological parameters found to be optimally constrained by the standard analysis with power spectra.
 
 \begin{table}
 \begin{center}
 \begin{tabular}{|l|c|c|}
 \hline 
 & Power spectra & {\sc BORG-WL} \\
 \hline 
 $\Omega_\mathrm{m}$ & $0.3^{+0.5}_{-0.2}$ & $0.34^{+0.07}_{-0.05}$\\
 \\
 $\sigma_8$ & $0.6^{+0.4}_{-0.3}$ & $0.79^{+0.11}_{-0.11}$ \\
 \\
 $S_8$ &  $0.80^{+0.15}_{-0.10}$ & $0.83^{+0.10}_{-0.10}$ \\
 \hline\\
 \end{tabular}
 \caption{Mean of the posterior, and $68.3\%$ credible intervals for $\Omega_\mathrm{m}$ and $\sigma_8$ from tomographic power spectra $C_\ell^{ab}$ and {\sc BORG-WL}.}
 \label{Table1}
 \end{center}
 \end{table}
 
 By building this Bayesian field-based approach, we can extract information from the data beyond the two-point statistics and lift the $\Omega_\mathrm{m}$-$\sigma_8$ degeneracy that results from using the standard approach in weak lensing analysis to date. This demonstrates that the high-order statistics, which are overlooked by the standard methods, contain significant cosmological information.

The spectacular improvement is in line with the results of \cite{LeclercqHeavens2021}, who showed for a simplified lognormal model that data assimilation approaches can lift degeneracies that are present in analyses based on two-point summary statistics, even for nearly Gaussian fields. 

\subsection{Efficiency of the cosmology sampler}
\label{sub:efficiency}

By construction, subsequent samples in Markov chains are correlated. The statistical efficiency of an MCMC is determined by the number of independent samples that can be drawn from a chain of a certain length. To estimate the efficiency of the sampler, we measure the correlation length of the cosmological parameters along the Markov chain. We typically  define the correlation length as the lag at which the auto-correlation drops below 0.1  \citep{Gregory2010} Figure \ref{fig:corr} presents the results of this test, showing that the correlation length for the parameters $\Omega_\mathrm{m}$ and $\sigma_8$ are 600 and 200 samples respectively. The parameter $h$ shows a very short correlation length because the data is not constraining.

\subsection{Convergence tests}
\label{sub:convergence}

To assess whether the Markov chain has fully explored the target posterior distribution or more samples are required, we consider the convergence of the Markov chain. A common approach to assess the convergence is based on comparing multiple chains with different starting points. One way to compare the chains is through a Gelman-Rubin diagnostic \citep{Gelman92}, where the convergence is assessed by comparing the intra-chains and inter-chains variances for each model parameter. If the difference between these variances is large, the chains have not converged.

In this work, we run two chains with starting points $(\Omega_\mathrm{m}, h, \sigma_8) = (0.31, 0.677, 0.8)$ and $(\Omega_\mathrm{m}, h, \sigma_8) = (0.4, 0.7, 0.55)$ respectively. The initial conditions are also randomised. After 3000 samples (discarding the burn-in phase), the Gelman-Rubin diagnostic is $R < 1.03$ for all the cosmological parameters. This indicates that the chains have converged reasonably well. Figure \ref{fig:convergence} presents the $\Omega_\mathrm{m}$-$\sigma_8$ constraints from both chains, showing that they converge to the same results. 

\subsection{Inferred density fields}
\label{sub:validation}

Jointly with the cosmology results presented in the previous sections, {\sc BORG-WL} also infers the 3D matter density field. Since we have shown that the method correctly recovers the underlying cosmology, here we focus on testing and validating the field results. To do so, we compare the inferred density fields to the truth and test whether the inferred cosmology and densities can explain the data. Although the tests presented here are the same as in our previous work \cite{BORG-WL}, there is a significant difference: here, we sample the cosmology and the density field jointly, while in our previous work, we used a fixed cosmology. Therefore, it is important to validate this extension of the method by running a similar set of consistency tests to confirm that {\sc BORG-WL} recovers the density fields when the cosmology is sampled.

Since {\sc BORG-WL} uses MCMC to explore the parameter space, we have access to the full posterior distribution, and we can draw posterior samples of the primordial matter fluctuations and the matter distribution. We use these samples to compute the mean and uncertainty of the inferred fields. Figure \ref{fig:panels_density} shows the projection on the sky of the true density fields and the mean and variance of the posterior samples. Since the lensing data is not very informative in the radial direction, we projected the three-dimensional fields on the sky, weighting the planes of the density field with the distribution of sources. A visual comparison between the true and inferred fields shows that the method correctly recovers the structures from the truth. As expected, the estimated mean of the fields shows some degree of smoothing, a feature familiar from Wiener filtering \citep{Gregory2010}.  This smoothing is an effect of averaging several samples, but the individual posterior samples have the correct amplitude and power spectrum. Figure \ref{fig:pk} shows the matter power spectrum of the initial conditions. The power spectrum is directly measured from the three-dimensional samples and agrees with the truth, indicating that the posterior samples have the correct clustering properties.

To test whether the inferred cosmology and densities provide accurate explanations of the data, we do a posterior predictive test \citep[][]{gelmanbda04}. This consists of applying the data model to the inferred quantities and checking whether we recover the structures in the data. Figure \ref{fig:panels_ppt} shows the results of this test for one tomographic bin, showing that the posterior-predicted shear field has the features of the true shear, and, therefore, the model can explain the data. At the resolution of this work, we are in the weakly non-Gaussian regime. We can expect a further improvement of the cosmology constraints when going to a higher resolution, but at smaller scales the baryon effects will be important. In a follow-up work, we will incorporate baryon feedback into the data model and increase the resolution of the analysis.

\section{Summary and conclusions}
\label{sec:conclusions}

We have presented a field-based approach to infer jointly cosmological parameters and the dark matter distribution, extending our earlier {\sc BORG-WL} implementation \citep{BORG-WL}, by sampling cosmological parameters and upgrading the line-of-sight integrations. The forward model consists of uniform priors for the cosmological parameters, a Gaussian prior for the primordial fluctuations, and a physical gravity model for the non-linear dynamics that link the initial conditions and the evolved density field. The cosmological parameters are sampled and updated consistently throughout the forward model by varying the initial matter power spectrum, the geometry of the lines of sight (through the distance-redshift relation) and the growth of cosmic structures (through the dynamical model and light-cone). This allows us to constrain the cosmology and the matter distribution simultaneously. 

We demonstrated the application of this method on simulated shear data with four tomographic bins and the number of sources per square-arcmin expected for {\it Euclid}. We sampled $\Omega_\mathrm{m}$, $h$ and $\sigma_8$ and compared the constraints from our field-based approach to the standard analysis based on the power spectra, with the same simulated data. The results showed that {\sc BORG-WL} lifts the $\Omega_\mathrm{m}$-$\sigma_8$ degeneracy and provides far more precise constraints from the same underlying data: the uncertainties in $\Omega_\mathrm{m}$ and $\sigma_8$ are decreased enormously, by a factor of 5 and 3, respectively. This demonstrates the advantage of using data assimilation techniques and applying the likelihood to the shear fields themselves, rather than to summary two-point statistics. 

In future work, we will extend our forward model to include systematic effects such as intrinsic alignments, baryon feedback, and uncertain redshift distributions, and apply this method to real data to constrain jointly cosmological parameters and the underlying matter distribution.   

\section*{Acknowledgements}
This work was supported by STFC through Imperial College Astrophysics Consolidated Grant ST/5000372/1. GL acknowledges financial support from the ANR BIG4, under reference ANR-16-CE23-0002. 
This work was carried out within the Aquila Consortium\footnote{\url{https://aquila-consortium.org}}.

%%%%%%%%%%%%%%%%%%%%%%%%%%%%%%%%%%%%%%%%%%%%%%%%%%
\section*{Data Availability}
 The data underlying this article will be shared on reasonable request to the corresponding author.



\bibliographystyle{mnras}
\bibliography{lensing} % if your bibtex file is called example.bib

\begin{thebibliography}{}
\makeatletter
\relax
\def\mn@urlcharsother{\let\do\@makeother \do\$\do\&\do\#\do\^\do\_\do\%\do\~}
\def\mn@doi{\begingroup\mn@urlcharsother \@ifnextchar [ {\mn@doi@}
  {\mn@doi@[]}}
\def\mn@doi@[#1]#2{\def\@tempa{#1}\ifx\@tempa\@empty \href
  {http://dx.doi.org/#2} {doi:#2}\else \href {http://dx.doi.org/#2} {#1}\fi
  \endgroup}
\def\mn@eprint#1#2{\mn@eprint@#1:#2::\@nil}
\def\mn@eprint@arXiv#1{\href {http://arxiv.org/abs/#1} {{\tt arXiv:#1}}}
\def\mn@eprint@dblp#1{\href {http://dblp.uni-trier.de/rec/bibtex/#1.xml}
  {dblp:#1}}
\def\mn@eprint@#1:#2:#3:#4\@nil{\def\@tempa {#1}\def\@tempb {#2}\def\@tempc
  {#3}\ifx \@tempc \@empty \let \@tempc \@tempb \let \@tempb \@tempa \fi \ifx
  \@tempb \@empty \def\@tempb {arXiv}\fi \@ifundefined
  {mn@eprint@\@tempb}{\@tempb:\@tempc}{\expandafter \expandafter \csname
  mn@eprint@\@tempb\endcsname \expandafter{\@tempc}}}

\bibitem[\protect\citeauthoryear{{Alsing}, {Heavens}, {Jaffe}, {Kiessling},
  {Wandelt}  \& {Hoffmann}}{{Alsing} et~al.}{2016}]{Alsing16}
{Alsing} J.,  {Heavens} A.,  {Jaffe} A.~H.,  {Kiessling} A.,  {Wandelt} B.,
  {Hoffmann} T.,  2016, \mn@doi [\mnras] {10.1093/mnras/stv2501}, \href
  {https://ui.adsabs.harvard.edu/abs/2016MNRAS.455.4452A} {455, 4452}

\bibitem[\protect\citeauthoryear{{Alsing}, {Heavens}  \& {Jaffe}}{{Alsing}
  et~al.}{2017}]{Alsing17}
{Alsing} J.,  {Heavens} A.,   {Jaffe} A.~H.,  2017, \mn@doi [\mnras]
  {10.1093/mnras/stw3161}, \href
  {https://ui.adsabs.harvard.edu/abs/2017MNRAS.466.3272A} {466, 3272}

\bibitem[\protect\citeauthoryear{{Amon} et~al.,}{{Amon}
  et~al.}{2021}]{Amon2021}
{Amon} A.,  et~al., 2021, arXiv e-prints, \href
  {https://ui.adsabs.harvard.edu/abs/2021arXiv210513543A} {p. arXiv:2105.13543}

\bibitem[\protect\citeauthoryear{{Asgari} et~al.,}{{Asgari}
  et~al.}{2021}]{Asgari2021}
{Asgari} M.,  et~al., 2021, \mn@doi [\aap] {10.1051/0004-6361/202039070}, \href
  {https://ui.adsabs.harvard.edu/abs/2021A&A...645A.104A} {645, A104}

\bibitem[\protect\citeauthoryear{{Bernardeau}}{{Bernardeau}}{2005}]{Bernardeau2005Third}
{Bernardeau} F.,  2005, \mn@doi [\aap] {10.1051/0004-6361:20053440}, \href
  {https://ui.adsabs.harvard.edu/abs/2005A&A...441..873B} {441, 873}

\bibitem[\protect\citeauthoryear{{B{\"o}hm}, {Hilbert}, {Greiner}  \&
  {En{\ss}lin}}{{B{\"o}hm} et~al.}{2017}]{Boehm17}
{B{\"o}hm} V.,  {Hilbert} S.,  {Greiner} M.,   {En{\ss}lin} T.~A.,  2017,
  \mn@doi [\prd] {10.1103/PhysRevD.96.123510}, \href
  {https://ui.adsabs.harvard.edu/abs/2017PhRvD..96l3510B} {96, 123510}

\bibitem[\protect\citeauthoryear{{Boyle}, {Uhlemann}, {Friedrich},
  {Barthelemy}, {Codis}, {Bernardeau}, {Giocoli}  \& {Baldi}}{{Boyle}
  et~al.}{2021}]{Boyle2021PDF}
{Boyle} A.,  {Uhlemann} C.,  {Friedrich} O.,  {Barthelemy} A.,  {Codis} S.,
  {Bernardeau} F.,  {Giocoli} C.,   {Baldi} M.,  2021, \mn@doi [\mnras]
  {10.1093/mnras/stab1381}, \href
  {https://ui.adsabs.harvard.edu/abs/2021MNRAS.505.2886B} {505, 2886}

\bibitem[\protect\citeauthoryear{{Chisari} et~al.,}{{Chisari}
  et~al.}{2019}]{CCL}
{Chisari} N.~E.,  et~al., 2019, \mn@doi [\apjs] {10.3847/1538-4365/ab1658},
  \href {https://ui.adsabs.harvard.edu/abs/2019ApJS..242....2C} {242, 2}

\bibitem[\protect\citeauthoryear{{Dietrich} \& {Hartlap}}{{Dietrich} \&
  {Hartlap}}{2010}]{Dietrich2010}
{Dietrich} J.~P.,  {Hartlap} J.,  2010, \mn@doi [\mnras]
  {10.1111/j.1365-2966.2009.15948.x}, \href
  {https://ui.adsabs.harvard.edu/abs/2010MNRAS.402.1049D} {402, 1049}

\bibitem[\protect\citeauthoryear{{Eisenstein} \& {Hu}}{{Eisenstein} \&
  {Hu}}{1998}]{EH98}
{Eisenstein} D.~J.,  {Hu} W.,  1998, \mn@doi [\apj] {10.1086/305424}, \href
  {http://adsabs.harvard.edu/abs/1998ApJ...496..605E} {496, 605}

\bibitem[\protect\citeauthoryear{{Eisenstein} \& {Hu}}{{Eisenstein} \&
  {Hu}}{1999}]{EH99}
{Eisenstein} D.~J.,  {Hu} W.,  1999, \mn@doi [\apj] {10.1086/306640}, \href
  {http://adsabs.harvard.edu/cgi-bin/nph-bib_query?bibcode=1999ApJ...511....5E&db_key=AST}
  {511, 5}

\bibitem[\protect\citeauthoryear{{Fiedorowicz}, {Rozo}, {Boruah}, {Chang}  \&
  {Gatti}}{{Fiedorowicz} et~al.}{2021}]{Fiedorowicz2021}
{Fiedorowicz} P.,  {Rozo} E.,  {Boruah} S.~S.,  {Chang} C.,   {Gatti} M.,
  2021, arXiv e-prints, \href
  {https://ui.adsabs.harvard.edu/abs/2021arXiv210514699F} {p. arXiv:2105.14699}

\bibitem[\protect\citeauthoryear{{Fluri}, {Kacprzak}, {Refregier}, {Amara},
  {Lucchi}  \& {Hofmann}}{{Fluri} et~al.}{2018a}]{Fluri2018ML}
{Fluri} J.,  {Kacprzak} T.,  {Refregier} A.,  {Amara} A.,  {Lucchi} A.,
  {Hofmann} T.,  2018a, \mn@doi [\prd] {10.1103/PhysRevD.98.123518}, \href
  {https://ui.adsabs.harvard.edu/abs/2018PhRvD..98l3518F} {98, 123518}

\bibitem[\protect\citeauthoryear{{Fluri}, {Kacprzak}, {Sgier}, {Refregier}  \&
  {Amara}}{{Fluri} et~al.}{2018b}]{Fluri2018Peaks}
{Fluri} J.,  {Kacprzak} T.,  {Sgier} R.,  {Refregier} A.,   {Amara} A.,  2018b,
  \mn@doi [\jcap] {10.1088/1475-7516/2018/10/051}, \href
  {https://ui.adsabs.harvard.edu/abs/2018JCAP...10..051F} {2018, 051}

\bibitem[\protect\citeauthoryear{{Fu} et~al.,}{{Fu} et~al.}{2014}]{Fu2014Third}
{Fu} L.,  et~al., 2014, \mn@doi [\mnras] {10.1093/mnras/stu754}, \href
  {https://ui.adsabs.harvard.edu/abs/2014MNRAS.441.2725F} {441, 2725}

\bibitem[\protect\citeauthoryear{Gelman \& Rubin}{Gelman \&
  Rubin}{1992}]{Gelman92}
Gelman A.,  Rubin D.~B.,  1992, \mn@doi [Statistical Science]
  {10.1214/ss/1177011136}, 7, 457

\bibitem[\protect\citeauthoryear{Gelman, Carlin, Stern  \& Rubin}{Gelman
  et~al.}{2004}]{gelmanbda04}
Gelman A.,  Carlin J.~B.,  Stern H.~S.,   Rubin D.~B.,  2004, Bayesian Data
  Analysis, 2nd ed. edn.
Chapman and Hall/CRC

\bibitem[\protect\citeauthoryear{{Giblin} et~al.,}{{Giblin}
  et~al.}{2018}]{Giblin2018}
{Giblin} B.,  et~al., 2018, \mn@doi [\mnras] {10.1093/mnras/sty2271}, \href
  {https://ui.adsabs.harvard.edu/abs/2018MNRAS.480.5529G} {480, 5529}

\bibitem[\protect\citeauthoryear{{Gregory}}{{Gregory}}{2010}]{Gregory2010}
{Gregory} P.,  2010, {Bayesian Logical Data Analysis for the Physical Sciences}

\bibitem[\protect\citeauthoryear{{Gupta}, {Matilla}, {Hsu}  \&
  {Haiman}}{{Gupta} et~al.}{2018}]{Gupta2018ML}
{Gupta} A.,  {Matilla} J. M.~Z.,  {Hsu} D.,   {Haiman} Z.,  2018, \mn@doi
  [\prd] {10.1103/PhysRevD.97.103515}, \href
  {https://ui.adsabs.harvard.edu/abs/2018PhRvD..97j3515G} {97, 103515}

\bibitem[\protect\citeauthoryear{{Hall}}{{Hall}}{2021}]{Hall2021}
{Hall} A.,  2021, \mn@doi [\mnras] {10.1093/mnras/stab1563}, \href
  {https://ui.adsabs.harvard.edu/abs/2021MNRAS.505.4935H} {505, 4935}

\bibitem[\protect\citeauthoryear{{Hamana} et~al.,}{{Hamana}
  et~al.}{2020}]{Hamana2020}
{Hamana} T.,  et~al., 2020, \mn@doi [\pasj] {10.1093/pasj/psz138}, \href
  {https://ui.adsabs.harvard.edu/abs/2020PASJ...72...16H} {72, 16}

\bibitem[\protect\citeauthoryear{{Harnois-D{\'e}raps}, {Martinet}, {Castro},
  {Dolag}, {Giblin}, {Heymans}, {Hildebrandt}  \& {Xia}}{{Harnois-D{\'e}raps}
  et~al.}{2021}]{Harnois2021Peaks}
{Harnois-D{\'e}raps} J.,  {Martinet} N.,  {Castro} T.,  {Dolag} K.,  {Giblin}
  B.,  {Heymans} C.,  {Hildebrandt} H.,   {Xia} Q.,  2021, \mn@doi [\mnras]
  {10.1093/mnras/stab1623}, \href
  {https://ui.adsabs.harvard.edu/abs/2021MNRAS.506.1623H} {506, 1623}

\bibitem[\protect\citeauthoryear{{Hikage} et~al.,}{{Hikage}
  et~al.}{2019}]{Hikage19}
{Hikage} C.,  et~al., 2019, \mn@doi [\pasj] {10.1093/pasj/psz010}, \href
  {https://ui.adsabs.harvard.edu/abs/2019PASJ...71...43H} {71, 43}

\bibitem[\protect\citeauthoryear{{Hildebrandt} et~al.,}{{Hildebrandt}
  et~al.}{2017}]{Hildebrandt2017}
{Hildebrandt} H.,  et~al., 2017, \mn@doi [\mnras] {10.1093/mnras/stw2805},
  \href {https://ui.adsabs.harvard.edu/abs/2017MNRAS.465.1454H} {465, 1454}

\bibitem[\protect\citeauthoryear{{Jain} \& {van Waerbeke}}{{Jain} \& {van
  Waerbeke}}{2000}]{Jain2000}
{Jain} B.,  {van Waerbeke} L.,  2000, \mn@doi [\apjl] {10.1086/312480}, \href
  {https://ui.adsabs.harvard.edu/abs/2000ApJ...530L...1J} {530, L1}

\bibitem[\protect\citeauthoryear{{Jarvis}, {Bernstein}  \& {Jain}}{{Jarvis}
  et~al.}{2004}]{Jarvis2004}
{Jarvis} M.,  {Bernstein} G.,   {Jain} B.,  2004, \mn@doi [\mnras]
  {10.1111/j.1365-2966.2004.07926.x}, \href
  {https://ui.adsabs.harvard.edu/abs/2004MNRAS.352..338J} {352, 338}

\bibitem[\protect\citeauthoryear{{Jasche} \& {Kitaura}}{{Jasche} \&
  {Kitaura}}{2010}]{HADES}
{Jasche} J.,  {Kitaura} F.~S.,  2010, \mn@doi [\mnras]
  {10.1111/j.1365-2966.2010.16897.x}, \href
  {http://adsabs.harvard.edu/abs/2010MNRAS.407...29J} {407, 29}

\bibitem[\protect\citeauthoryear{{Jasche} \& {Lavaux}}{{Jasche} \&
  {Lavaux}}{2019}]{Jasche18BorgPM}
{Jasche} J.,  {Lavaux} G.,  2019, \mn@doi [\aap] {10.1051/0004-6361/201833710},
  \href {https://ui.adsabs.harvard.edu/abs/2019A&A...625A..64J} {625, A64}

\bibitem[\protect\citeauthoryear{{Jasche} \& {Wandelt}}{{Jasche} \&
  {Wandelt}}{2013}]{BORG}
{Jasche} J.,  {Wandelt} B.~D.,  2013, \mn@doi [\mnras] {10.1093/mnras/stt449},
  \href {http://adsabs.harvard.edu/abs/2013MNRAS.432..894J} {432, 894}

\bibitem[\protect\citeauthoryear{{Jeffrey}, {Alsing}  \& {Lanusse}}{{Jeffrey}
  et~al.}{2021}]{Jeffrey2021ML}
{Jeffrey} N.,  {Alsing} J.,   {Lanusse} F.,  2021, \mn@doi [\mnras]
  {10.1093/mnras/staa3594}, \href
  {https://ui.adsabs.harvard.edu/abs/2021MNRAS.501..954J} {501, 954}

\bibitem[\protect\citeauthoryear{{Jung}, {Namikawa}, {Liguori}, {Munshi}  \&
  {Heavens}}{{Jung} et~al.}{2021}]{Jung2021}
{Jung} G.,  {Namikawa} T.,  {Liguori} M.,  {Munshi} D.,   {Heavens} A.,  2021,
  \mn@doi [\jcap] {10.1088/1475-7516/2021/06/055}, \href
  {https://ui.adsabs.harvard.edu/abs/2021JCAP...06..055J} {2021, 055}

\bibitem[\protect\citeauthoryear{{Kacprzak} et~al.,}{{Kacprzak}
  et~al.}{2016}]{Kacprzak2016Peaks}
{Kacprzak} T.,  et~al., 2016, \mn@doi [\mnras] {10.1093/mnras/stw2070}, \href
  {https://ui.adsabs.harvard.edu/abs/2016MNRAS.463.3653K} {463, 3653}

\bibitem[\protect\citeauthoryear{{Kilbinger}}{{Kilbinger}}{2015}]{KilbingerReview}
{Kilbinger} M.,  2015, \mn@doi [Reports on Progress in Physics]
  {10.1088/0034-4885/78/8/086901}, \href
  {https://ui.adsabs.harvard.edu/abs/2015RPPh...78h6901K} {78, 086901}

\bibitem[\protect\citeauthoryear{{Kilbinger} \& {Schneider}}{{Kilbinger} \&
  {Schneider}}{2005}]{Kilbinger2005Third}
{Kilbinger} M.,  {Schneider} P.,  2005, \mn@doi [\aap]
  {10.1051/0004-6361:20053531}, \href
  {https://ui.adsabs.harvard.edu/abs/2005A&A...442...69K} {442, 69}

\bibitem[\protect\citeauthoryear{{Laureijs} et~al.,}{{Laureijs}
  et~al.}{2011}]{EuclidStudyReport}
{Laureijs} R.,  et~al., 2011, arXiv e-prints, \href
  {https://ui.adsabs.harvard.edu/abs/2011arXiv1110.3193L} {p. arXiv:1110.3193}

\bibitem[\protect\citeauthoryear{{Lavaux}, {Jasche}  \& {Leclercq}}{{Lavaux}
  et~al.}{2019}]{BORG-3}
{Lavaux} G.,  {Jasche} J.,   {Leclercq} F.,  2019, arXiv e-prints, \href
  {https://ui.adsabs.harvard.edu/abs/2019arXiv190906396L} {p. arXiv:1909.06396}

\bibitem[\protect\citeauthoryear{{Leclercq} \& {Heavens}}{{Leclercq} \&
  {Heavens}}{2021}]{LeclercqHeavens2021}
{Leclercq} F.,  {Heavens} A.,  2021, arXiv e-prints, \href
  {https://ui.adsabs.harvard.edu/abs/2021arXiv210304158L} {p. arXiv:2103.04158}

\bibitem[\protect\citeauthoryear{{Lin} \& {Kilbinger}}{{Lin} \&
  {Kilbinger}}{2015}]{Lin2015}
{Lin} C.-A.,  {Kilbinger} M.,  2015, \mn@doi [\aap]
  {10.1051/0004-6361/201526659}, \href
  {https://ui.adsabs.harvard.edu/abs/2015A&A...583A..70L} {583, A70}

\bibitem[\protect\citeauthoryear{{Liu}, {Petri}, {Haiman}, {Hui}, {Kratochvil}
  \& {May}}{{Liu} et~al.}{2015}]{Liu2015Peaks}
{Liu} J.,  {Petri} A.,  {Haiman} Z.,  {Hui} L.,  {Kratochvil} J.~M.,   {May}
  M.,  2015, \mn@doi [\prd] {10.1103/PhysRevD.91.063507}, \href
  {https://ui.adsabs.harvard.edu/abs/2015PhRvD..91f3507L} {91, 063507}

\bibitem[\protect\citeauthoryear{{Martinet} et~al.,}{{Martinet}
  et~al.}{2018}]{Martinet2018Peaks}
{Martinet} N.,  et~al., 2018, \mn@doi [\mnras] {10.1093/mnras/stx2793}, \href
  {https://ui.adsabs.harvard.edu/abs/2018MNRAS.474..712M} {474, 712}

\bibitem[\protect\citeauthoryear{{Martinet}, {Harnois-D{\'e}raps}, {Jullo}  \&
  {Schneider}}{{Martinet} et~al.}{2021}]{Martinet2021PDF}
{Martinet} N.,  {Harnois-D{\'e}raps} J.,  {Jullo} E.,   {Schneider} P.,  2021,
  \mn@doi [\aap] {10.1051/0004-6361/202039679}, \href
  {https://ui.adsabs.harvard.edu/abs/2021A&A...646A..62M} {646, A62}

\bibitem[\protect\citeauthoryear{{Maturi}, {Fedeli}  \& {Moscardini}}{{Maturi}
  et~al.}{2011}]{Maturi2011}
{Maturi} M.,  {Fedeli} C.,   {Moscardini} L.,  2011, \mn@doi [\mnras]
  {10.1111/j.1365-2966.2011.18958.x}, \href
  {https://ui.adsabs.harvard.edu/abs/2011MNRAS.416.2527M} {416, 2527}

\bibitem[\protect\citeauthoryear{{Neal}}{{Neal}}{2011}]{Neal2011}
{Neal} R.,  2011, {MCMC Using Hamiltonian Dynamics}.
pp 113--162, \mn@doi{10.1201/b10905}

\bibitem[\protect\citeauthoryear{{Peel}, {Lin}, {Lanusse}, {Leonard}, {Starck}
  \& {Kilbinger}}{{Peel} et~al.}{2017}]{Peel2017}
{Peel} A.,  {Lin} C.-A.,  {Lanusse} F.,  {Leonard} A.,  {Starck} J.-L.,
  {Kilbinger} M.,  2017, \mn@doi [\aap] {10.1051/0004-6361/201629928}, \href
  {https://ui.adsabs.harvard.edu/abs/2017A&A...599A..79P} {599, A79}

\bibitem[\protect\citeauthoryear{{Pen}, {Zhang}, {van Waerbeke}, {Mellier},
  {Zhang}  \& {Dubinski}}{{Pen} et~al.}{2003}]{Pen2003}
{Pen} U.-L.,  {Zhang} T.,  {van Waerbeke} L.,  {Mellier} Y.,  {Zhang} P.,
  {Dubinski} J.,  2003, \mn@doi [\apj] {10.1086/375734}, \href
  {https://ui.adsabs.harvard.edu/abs/2003ApJ...592..664P} {592, 664}

\bibitem[\protect\citeauthoryear{{Petri}, {Haiman}, {Hui}, {May}  \&
  {Kratochvil}}{{Petri} et~al.}{2013}]{Petri2013}
{Petri} A.,  {Haiman} Z.,  {Hui} L.,  {May} M.,   {Kratochvil} J.~M.,  2013,
  \mn@doi [\prd] {10.1103/PhysRevD.88.123002}, \href
  {https://ui.adsabs.harvard.edu/abs/2013PhRvD..88l3002P} {88, 123002}

\bibitem[\protect\citeauthoryear{{Porqueres}, {Heavens}, {Mortlock}  \&
  {Lavaux}}{{Porqueres} et~al.}{2021}]{BORG-WL}
{Porqueres} N.,  {Heavens} A.,  {Mortlock} D.,   {Lavaux} G.,  2021, \mn@doi
  [\mnras] {10.1093/mnras/stab204}, \href
  {https://ui.adsabs.harvard.edu/abs/2021MNRAS.502.3035P} {502, 3035}

\bibitem[\protect\citeauthoryear{{Ramanah}, {Lavaux}, {Jasche}  \& {Wand
  elt}}{{Ramanah} et~al.}{2019}]{Altair}
{Ramanah} D.~K.,  {Lavaux} G.,  {Jasche} J.,   {Wand elt} B.~D.,  2019, \mn@doi
  [\aap] {10.1051/0004-6361/201834117}, \href
  {https://ui.adsabs.harvard.edu/abs/2019A&A...621A..69R} {621, A69}

\bibitem[\protect\citeauthoryear{{Ribli}, {Pataki}, {Zorrilla Matilla}, {Hsu},
  {Haiman}  \& {Csabai}}{{Ribli} et~al.}{2019}]{Ribli2019ML}
{Ribli} D.,  {Pataki} B.~{\'A}.,  {Zorrilla Matilla} J.~M.,  {Hsu} D.,
  {Haiman} Z.,   {Csabai} I.,  2019, \mn@doi [\mnras] {10.1093/mnras/stz2610},
  \href {https://ui.adsabs.harvard.edu/abs/2019MNRAS.490.1843R} {490, 1843}

\bibitem[\protect\citeauthoryear{{Schneider}, {Hogg}, {Marshall}, {Dawson},
  {Meyers}, {Bard}  \& {Lang}}{{Schneider} et~al.}{2015}]{Schneider15}
{Schneider} M.~D.,  {Hogg} D.~W.,  {Marshall} P.~J.,  {Dawson} W.~A.,  {Meyers}
  J.,  {Bard} D.~J.,   {Lang} D.,  2015, \mn@doi [\apj]
  {10.1088/0004-637X/807/1/87}, \href
  {https://ui.adsabs.harvard.edu/abs/2015ApJ...807...87S} {807, 87}

\bibitem[\protect\citeauthoryear{{Secco} et~al.,}{{Secco}
  et~al.}{2021}]{Secco2021}
{Secco} L.~F.,  et~al., 2021, arXiv e-prints, \href
  {https://ui.adsabs.harvard.edu/abs/2021arXiv210513544S} {p. arXiv:2105.13544}

\bibitem[\protect\citeauthoryear{{Semboloni}, {Schrabback}, {van Waerbeke},
  {Vafaei}, {Hartlap}  \& {Hilbert}}{{Semboloni}
  et~al.}{2011}]{Semboloni2011Third}
{Semboloni} E.,  {Schrabback} T.,  {van Waerbeke} L.,  {Vafaei} S.,  {Hartlap}
  J.,   {Hilbert} S.,  2011, \mn@doi [\mnras]
  {10.1111/j.1365-2966.2010.17430.x}, \href
  {https://ui.adsabs.harvard.edu/abs/2011MNRAS.410..143S} {410, 143}

\bibitem[\protect\citeauthoryear{{Shan} et~al.,}{{Shan}
  et~al.}{2018}]{Shan2018Peaks}
{Shan} H.,  et~al., 2018, \mn@doi [\mnras] {10.1093/mnras/stx2837}, \href
  {https://ui.adsabs.harvard.edu/abs/2018MNRAS.474.1116S} {474, 1116}

\bibitem[\protect\citeauthoryear{{Takada} \& {Jain}}{{Takada} \&
  {Jain}}{2002}]{Takada2002Third}
{Takada} M.,  {Jain} B.,  2002, \mn@doi [\mnras]
  {10.1046/j.1365-8711.2002.05972.x}, \href
  {https://ui.adsabs.harvard.edu/abs/2002MNRAS.337..875T} {337, 875}

\bibitem[\protect\citeauthoryear{{Tassev}, {Zaldarriaga}  \&
  {Eisenstein}}{{Tassev} et~al.}{2013}]{Tassev2013}
{Tassev} S.,  {Zaldarriaga} M.,   {Eisenstein} D.~J.,  2013, \mn@doi [\jcap]
  {10.1088/1475-7516/2013/06/036}, \href
  {https://ui.adsabs.harvard.edu/abs/2013JCAP...06..036T} {2013, 036}

\bibitem[\protect\citeauthoryear{{Troxel} et~al.,}{{Troxel}
  et~al.}{2018}]{Troxel2018}
{Troxel} M.~A.,  et~al., 2018, \mn@doi [\prd] {10.1103/PhysRevD.98.043528},
  \href {https://ui.adsabs.harvard.edu/abs/2018PhRvD..98d3528T} {98, 043528}

\bibitem[\protect\citeauthoryear{{van Waerbeke} et~al.,}{{van Waerbeke}
  et~al.}{2013}]{Waerbeke2013}
{van Waerbeke} L.,  et~al., 2013, \mn@doi [\mnras] {10.1093/mnras/stt971},
  \href {https://ui.adsabs.harvard.edu/abs/2013MNRAS.433.3373V} {433, 3373}

\makeatother
\end{thebibliography}

% Don't change these lines
\bsp	% typesetting comment
\label{lastpage}
\end{document}